\documentclass[aps,prb,groupedaddress,twocolumn,nofootinbib]{revtex4}
\usepackage{graphicx}
\usepackage{hyperref}
\usepackage[english]{babel}
\usepackage{makeidx}
\usepackage{amsfonts}
\usepackage{amssymb}
\usepackage{bm}
\begin{document}

\title{Prediction of the thermophysical properties of molten salt fast reactor fuel from first-principles}
\author{A. E. Gheribi$^a$, D. Corradini$^{b,c}$, L. Dewan$^d$, P. Chartrand$^{a}$, C. Simon$^{b,c}$, P. A. Madden$^e$ and M. Salanne$^{b,c}$}
\affiliation{$^a$ Centre for Research in Computational Thermochemistry, Department of Chemical Engineering, Ecole Polytechnique, C.P. 6079, Succursale ``Downtown'', Montreal (Quebec), Canada H3C 3A7}
\affiliation{$^b$ Sorbonne Universit\'es, UPMC Univ Paris 06, UMR 8234, PHENIX, F-75005, Paris, France}
\affiliation{$^c$ CNRS, UMR 8234, PHENIX, F-75005, Paris, France}
\affiliation{$^d$ Massachusetts Institute of Technology, Department of Nuclear Science and Engineering, 77 Massachusetts Avenue, Cambridge, MA 02139, USA}
\affiliation{$^e$ Department of Materials, University of Oxford, Parks Road, Oxford OX1 3PH, UK}

\begin{abstract}
Molten fluorides are known to show favorable thermophysical properties which make them good candidate coolants for nuclear fission reactors. Here we investigate the special case of  mixtures of lithium fluoride and thorium fluoride, which act both as coolant and fuel in the molten salt fast reactor concept. By using {\it ab initio} parameterized polarizable force fields, we show that it is possible to calculate the whole set of properties (density, thermal expansion, heat capacity, viscosity and thermal conductivity) which are necessary for assessing the heat transfer performance of the melt over the whole range of compositions and temperatures. We then deduce from our calculations several figures of merit which are important in helping the optimization of the design of molten salt fast reactors.
\end{abstract}

\maketitle

\section{Introduction}
The molten salt fast reactor (MSFR) is currently the most advanced concept of nuclear fission reactor involving the use of molten salts both as coolant and as fuel~\cite{locatelli2013a}. A great deal of research has recently been devoted to the optimization of its neutronic~\cite{mathieu2006a,aufiero2014a}, safety~\cite{brovchenko2013a}, recycling~\cite{delpech2009a,hamel2007a,gibilaro2009a,taxil2009a,gibilaro2011a} and thermodynamic~\cite{benes2009b,benes2013a,capelli2013a} characteristics. Its advantages are an intrinsic safety of the core, due to very negative feedback coefficients, a high working temperature which ensures a good efficiency, and the possibility to perform online refuelling and on-site treatment. But of particular interest is its high compositional flexibility. In particular, our precognized liquid fuel is a mixture of lithium fluoride and thorium fluoride (LiF-ThF$_4$). Being able to develop fission reactors based on the thorium cycle ensures a good availability of the resources for many countries. From the point of view of basic research, a consequence of the MSFR compositional flexibility is the need to generate substantial databases providing the variation of the properties of the melt with composition and temperature.

Due to the high temperature and to the corrosive character of molten fluorides, measuring their thermophysical properties is not a straightforward task, and the experimental results are generaly limited to pure salts or binary mixtures at the eutectic composition~\cite{benes2012a,benes2012b,khoklov2011a}. An efficient alternative is to use computer simulations. In particular, molecular dynamics (MD) simulations are able to provide all the necessary properties, such as the density, the heat capacity, the viscosity and the thermal conductivity. The necessary ingredients are a good model for the ionic interactions, and its associated parameters. In recent years, we have developed from first-principles, i.e. without involving any empirical information during the parameterization procedure, such interaction potentials for molten fluorides, in the framework of the polarizable ion model~\cite{madden2006a,salanne2011c,salanne2012b}. They were tested against many physical properties, as well as on diffraction and spectroscopy experiments. An excellent agreement was obtained for the full set of data~\cite{heaton2006a,pauvert2010a,levesque2013b}. In particular, for LiF-ThF$_4$ eutectic mixture, which corresponds to a ThF$_4$ mole fraction $x$~=~0.22, the largest deviation from experimental data observed was 5~\% for the density~\cite{dewan2013a}. Such a good predictive power from MD simulations has not been obtained yet for water, for example. It is thus possible to perform simulations under a large variety of thermodynamic conditions in order to feed the databases. In this work, we focus on the LiF-ThF$_4$ binary system, for which we calculate all the relevant physical properties over the whole range of compositions and temperatures. This allows us to map the variations of the thermophysical figures of merit which are necessary for the engineering of MSFRs.

\section{Model and methods}

The pair additive form of the interionic potential includes charge-charge, charge-dipole,
dipole-charge, dipole-dipole, repulsion, and two dispersion interaction terms, as follows~:
    \begin{eqnarray}
      U_{ij}\left( {\bf r}_{ij} \right) &=& \frac{q_{i}q_{j}} {r_{ij}}
      + \frac{ q_{i} {\bf r}_{ij} \cdot {\boldsymbol{\mu}}_{j} } {r_{ij}^3} f^{ij}_{4}\left( r_{ij} \right)
      - \frac{ {\boldsymbol{\mu} }_{i} \cdot {\bf r}_{ij} q_{j} } {r_{ij}^3} f^{ji}_{4}\left( r_{ij} \right) \nonumber \\
      & &+ \frac{ {\boldsymbol{\mu}}_{i} \cdot {\boldsymbol{\mu}}_{j} } {r_{ij}^3}
      - \frac{ 3 ( {\bf r}_{ij} \cdot {\boldsymbol{\mu}}_{i} ) 
               ( {\bf r}_{ij} \cdot {\boldsymbol{\mu}}_{j} ) } {r_{ij}^5} \nonumber \\
      & &+ B_{ij} \exp \left( - \alpha_{ij} r_{ij} \right)
      - \frac{C^{ij}_6} {r_{ij}^6} f^{ij}_{6}\left( r_{ij} \right) - \frac{C^{ij}_8} {r_{ij}^8} f^{ij}_{8}\left( r_{ij} \right)
     \end{eqnarray}
\noindent where $q_i$ and ${\boldsymbol{\mu}}_i$ are the charge and dipole moment of particle $i$, respectively,
and $f_{n}^{ij}$ the damping function for short-range correction of interactions
between charge and dipole and dispersion interactions~\cite{tang1984a}:
\begin{equation}
f^{ij}_n(r^{ij}) = 1-c_n^{ij}{\rm e}^{-b_n^{ij}\times r^{ij}}\sum_{k=0}^n \frac{(b_n^{ij}\times r^{ij})^k}{k!}.
\end{equation}
\noindent $\{B_{ij}, \alpha_{ij}, C^{ij}_6, C^{ij}_8, b_n^{ij}, c_n^{ij}\}$ are a set of parameters which were determined from first-principles calculations in our previous works~\cite{dewan2013a}. The dipole moment induced on each ion is a function of the polarizability and electric field
on the ion caused by the charges and the dipole moments of all the other ions.
The instantaneous dipole moment is determined self-consistently
at every time step  by minimization of the total energy using the conjugate gradient method.
The  charge-charge,
charge-dipole, and dipole-dipole contributions to the potential energy and forces on each ion are evaluated under
 periodic boundary conditions by using the Ewald summation technique~\cite{aguado2003a}. 

A first series of MD simulations was performed in the $NPT$ ensemble, where $N$ is the particle number
included in the simulation cell, $P$ is the pressure which was always set to 1~atm and $T$ is the temperature. The barostat and thermostat relaxation times were both set to 10~ps. This allowed us to determine the equilibrium density at a given thermodynamic point. Then a second series of simulations was performed at fixed density, i.e. in the $NVT$ ensemble where $V$ is the volume. The same relaxation time was used for the thermostat. The simulation timestep was 0.5~fs, and the total simulation time was 0.25~ns and 7.5~ns for the $NPT$ and $NVT$ simulations, respectively. All the calculation conditions are summarized in the supplementary information; in total the system was simulated at 62 thermodynamic points with this procedure.

\section{Results and discussion}

\begin{center}
\begin{table}
\begin{center}
    \begin{tabular} { ccccccccc }
\hline
FOM & Convection & R\'egime & $r$ &$s$ &$t$ & $u$ & $v$ & $w$ \\
\hline
1 & Forced & Turbulent & 1.00 & 0.2 & 0.0 & 2.0 & 2.8 & 0.0 \\
2 & Natural & Turbulent & 0.36 &0.2 & 1.0 & 2.0 & 1.8 & 0.0 \\
3 & Natural & Laminar &  0.50 & 1.0 & 1.0 & 2.0 & 1.0 & 0.0 \\
4 &   --    &    --   &  1.00 & 0.2 & 1.0 & 0.3 & 0.6 & 0.6 \\
\hline
\end{tabular}
  \caption{\label{tab:fom} Coefficients for the figures of merit used to evaluate the heat-transfer properties of molten fluorides~\cite{bonilla1958a}. FOM4 is used for the heat-exchanger area.}
\end{center}
\end{table}
\end{center}

Our {\it ab initio}-parameterized interaction potentials have been validated against many experimental data for similar systems. More precisely, they are able to reproduce with very high accuracy the measured EXAFS spectra~\cite{pauvert2010a} (LiF-ZrF$_4$, NaF-ZrF$_4$ and KF-ZrF$_4$), the Raman spectra~\cite{heaton2008a} (LiF-BeF$_2$), the X-ray diffraction~\cite{salanne2006a} (LiF-BeF$_2$), the electrical conductivity~\cite{levesque2013b} (LiF-NaF-ZrF$_4$ and LiF-YF$_3$), the diffusion coefficients~\cite{saroukanian2009a,levesque2013b} (LiF-KF and LiF-YF$_3$), the heat capacity~\cite{benes2012b} (CsF) and the viscosity~\cite{salanne2006a} (LiF-BeF$_2$). For the thermal conductivity, this quantity is very difficult to measure experimentally and our simulations have shown an excellent agreement in the case of molten chlorides~\cite{ohtori2009a}, for which accurate measurements could be made by using a method based on forced Rayleigh scattering. In the case of LiF-ThF$_4$, in our previous work~\cite{dewan2013a} we have evaluated the density, the viscosity and the electrical conductivity at the eutectic composition. Our  MD results showed a very good agreement with the corresponding experimental data~\cite{benes2012a}. In the present work, we can therefore consider our simulations to provide us accurate predictions.

In order to compare the performances of a series of coolants, it is useful to define
generalized heat-transfer metrics. Several figure of merits
(FOM) have been proposed by Bonilla to evaluate the
properties of molten fluorides as coolants for nuclear reactors~\cite{bonilla1958a}. They take the following general form:
\begin{equation}
{\rm FOM}=\left(\frac{\eta^s}{\beta^t \rho^u C_{p,m}^v\lambda^w}\right)^r
\label{eq:fomeq}
\end{equation}
\noindent where $\eta$ is the viscosity, $\beta$ the thermal expansion, $\rho$ the density, $C_{p,m}$ the massic heat capacity at fixed pressure and $\lambda$ the thermal conductivity of a given melt. The set of exponents $r$, $s$, $t$, $u$, $v$ and $w$ vary depending on the conditions. The corresponding values are summarized in Table \ref{tab:fom}.

\begin{figure}[ht!]
\begin{center}
 \includegraphics[width=\columnwidth]{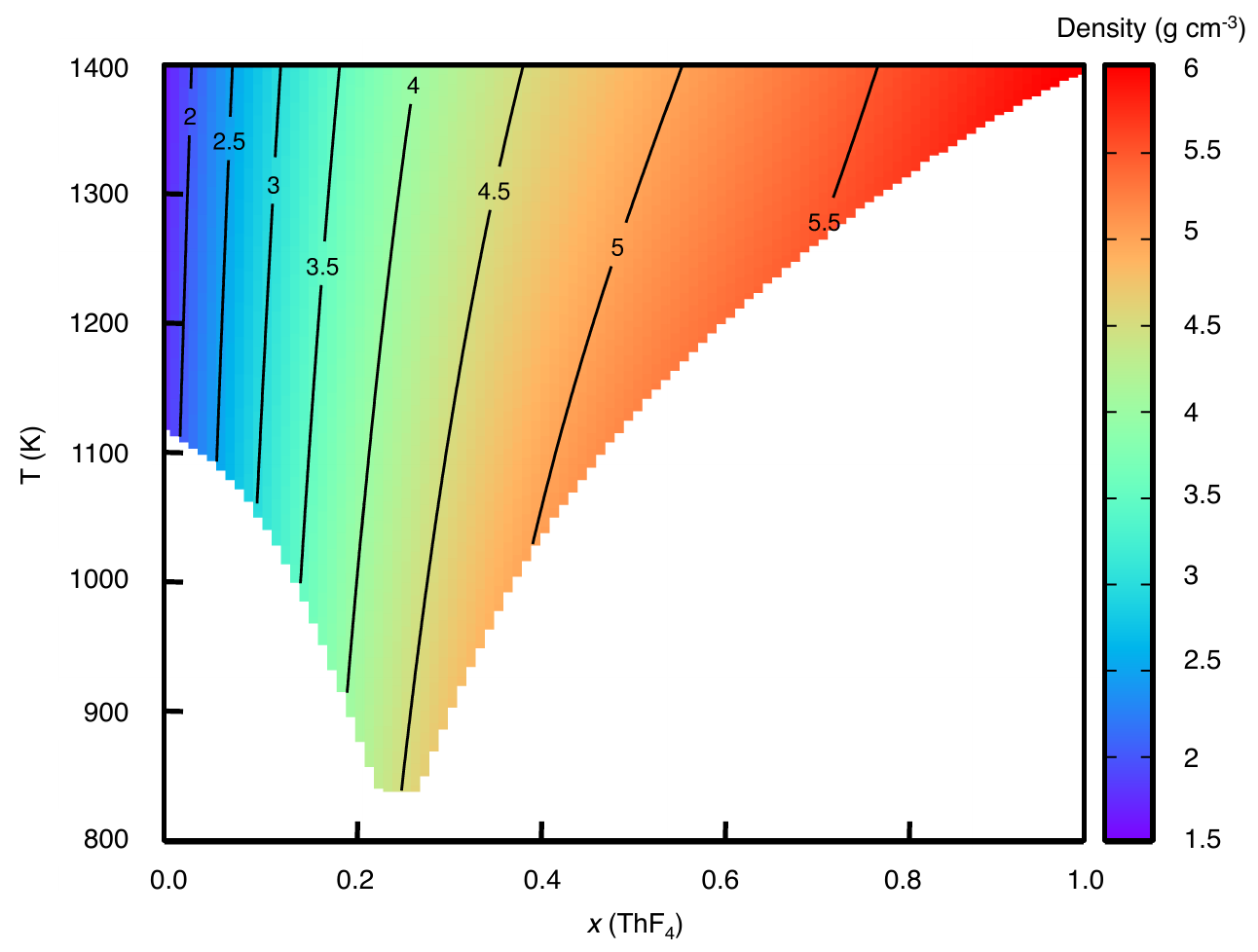}
\end{center}
\caption{Variation of the density of the LiF-ThF$_4$ melts with composition and temperature. The white region corresponds to temperatures below the liquidus.}
\label{fig:densite}
\end{figure}

Our objective is to map the variations of all the properties which are necessary for determining the FOMs over the whole range of temperatures and compositions. We have therefore tried to fit all the simulation data with simple analytic functions, by finding the parameters which allow to minimize
\begin{equation}
\chi^2(a) = \frac{1}{N_{\rm sim}}\sum_{i=1}^{N_{\rm sim}}\left(\frac{a_i(x,T)-a_i^{\rm MD}}{a_i^{\rm MD}}\right)^2
\end{equation}
\noindent where $a_i^{\rm MD}$ is the value extracted from the MD simulation for the property $a$ at a given thermodynamic point ($N_{\rm sim}$ is the total number of simulated points) and $a_i(x,T)$ is the corresponding value for the analytical function ($x$ is the mole fraction of ThF$_4$ in the melt and $T$ the temperature), which contains the parameters which need to be fitted.

First we start with the density. For this quantity, a small temperature dependence has to be included since this quantity usually varies linearly with temperature at fixed composition~\cite{benes2012a,dewan2013a}. An excellent fit of the simulation data was obtained by using the following polynomial expression:
\begin{eqnarray} 
  \rho(x,T)&=&2.2247-0.0004329\times T \nonumber \\
           & &+(15.412-0.0015748\times T)\times x  \nonumber\\
           & &+(-8.1106-0.0088374\times T)\times x^2 \nonumber \\
           & &+(-11.380+0.020484\times T)\times x^3 \nonumber\\
           & &+(9.2358-0.010673\times T)\times x^4
\end{eqnarray}
\noindent  The average error is $\chi(\rho)$~=~1.5~\%. The $\rho(x,T)$ map is plotted on Figure \ref{fig:densite}; unsuprisingly the largest variations are due to the composition changes, and larger densities are obtained for high ThF$_4$ concentrations due to the large difference of molar weight between the two cations.

The thermal expansion, which corresponds to the material's
volume change as a function of temperature, was derived
from the simulation results using the following equation:
\begin{equation}
\beta=-\frac{1}{\rho}\left(\frac{\partial \rho}{\partial T}\right)_P
\end{equation}
\noindent Its variation with temperature at fixed composition is negligible, and the following polynomial relation was fitted for the composition changes:
\begin{equation}
\beta(x,T)=\beta(x)=0.0001\times (3.0829-1.3187\times x-0.43862\times x^2)
\end{equation}
\noindent yielding a value of 0.8~\% for $\chi(\beta)$.

Similarly, the heat capacity of molten salts is usually constant with respect to temperature variations at fixed composition. This is not the case for the corresponding solid phases~\cite{benes2012b}. Our simulations, in which the molar heat capacity $C_p$ is calculated by differentiation of the molar enthalpy $H$ (extracted from the $NPT$ ensemble simulations) with respect to the temperature, 
\begin{equation}
C_p=\left(\frac{\partial H}{\partial T}\right)_P,
\end{equation}
\noindent confirm this behavior. Unlike other molten fluoride mixtures~\cite{beilmann2013a}, for which a deviation from ideality was measured, in the LiF-ThF$_4$ melts $C_p$ varies almost linearly with composition. However, a fourth degree polynomial analytic function was fitted,
\begin{eqnarray}
C_p(x,T)&=&C_p(x)=63.752+120.85\times x-115.22\times x^2 \nonumber \\
        & &+171.86\times x^3-68.994\times x^4
\end{eqnarray}
\noindent yielding 0.1~\% for $\chi(C_p)$. In order to deduce the massic heat capacity, we need to divide $C_p$ by the molar weight $M$, which is given exactly by the relation
\begin{equation}
M(x)=(232.04+18.998\times 4.0)\times x+(6.94+18.998)\times (1-x)
\end{equation}

\begin{figure}[ht!]
\begin{center}
 \includegraphics[width=\columnwidth]{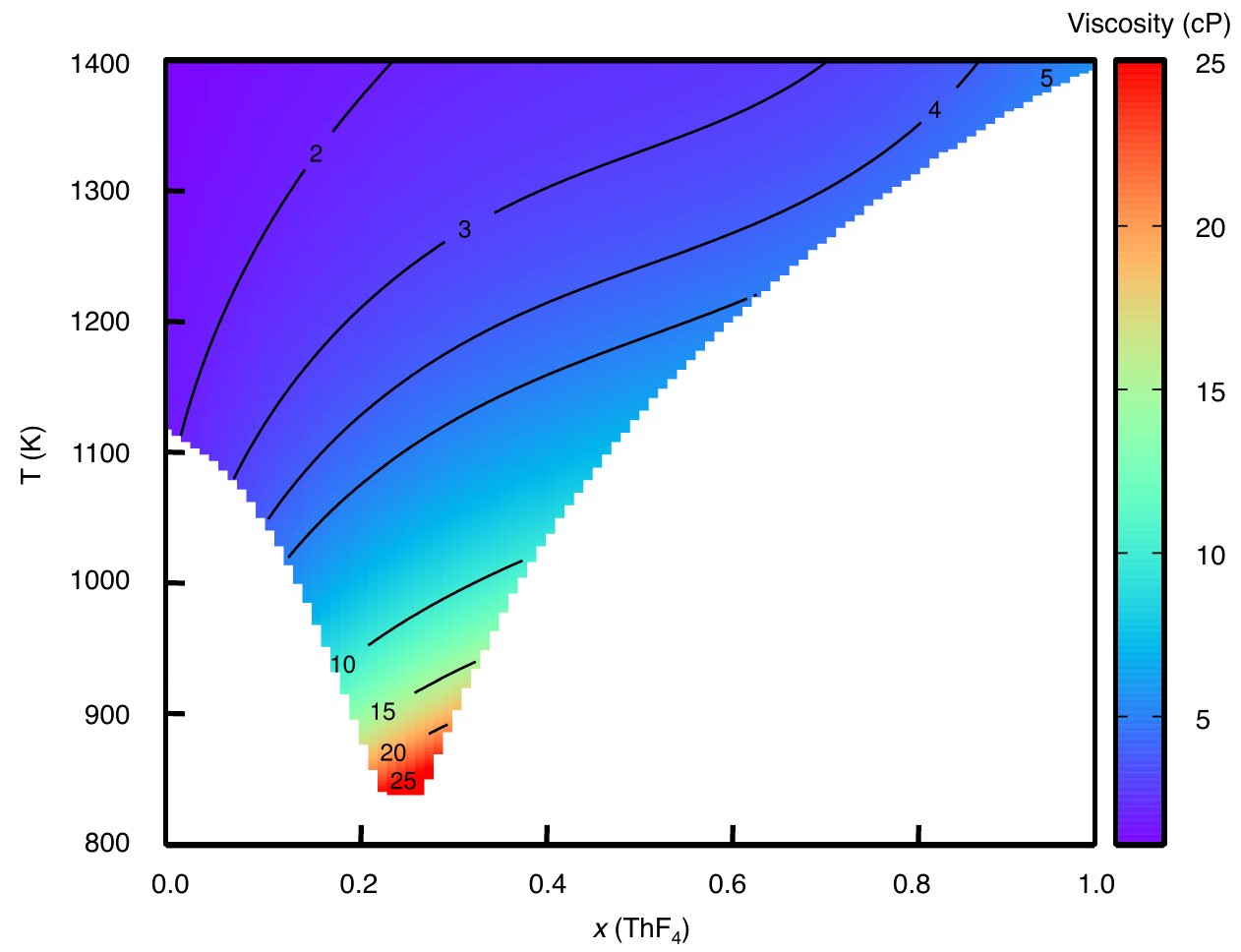}
\end{center}
\caption{Variation of the viscosity of the LiF-ThF$_4$ melts with composition and temperature.  The white region corresponds to temperatures below the liquidus.}
\label{fig:visco}
\end{figure}

The viscosity $\eta$ was determined by the Green-Kubo method, i.e. by 
integrating the correlation function of the stress tensor, as follows~:

\begin{equation}
\eta = \frac{V}{k_BT}\int_0^\infty \langle \sigma_{\alpha\beta}(0)\sigma_{\alpha\beta}(t)\rangle{\rm d}t
\label{eq:greenkubo}
\end{equation}

\noindent in which $\sigma_{\alpha\beta}$ is one of the components of the stress tensor~\cite{hansen-livre}. In order to improve the statistics, the stress autocorrelation function is averaged over the five independent components ($\sigma_{xy}$, $\sigma_{xz}$, $\sigma_{yz}$, $\sigma_{xx-yy}$ and $\sigma_{2zz-xx-yy}$). Much longer simulation times are needed to compute the viscosity, because enough statistics needs to be accumulated in order to yield a plateau for the integral defined in Equation \ref{eq:greenkubo}.

In molten salts, the viscosity typically follows an Arrhenius law for the variation with temperature. We have therefore used a more complex analytic form than for the other properties, 
\begin{eqnarray}
\eta(x,T)&=&0.21412\times \exp{\frac{2383.3}{T}} \nonumber \\
         & &+(0.022784\times \exp{\frac{7000.1}{T}})\times x \nonumber \\
         & & +(94.189\times \exp{\frac{-4807.0}{T}})\times x^2 \nonumber \\
         & &+(-3.0284\times \exp{\frac{2281.7}{T}})\times x^3 \nonumber \\
         & &+(0.76363\times \exp{\frac{4016.4}{T}})\times x^4
\end{eqnarray}
\noindent Although a larger overall error is observed, $\chi(\eta)$=9~\%, it remains within the error bars of the MD simulation values. The variation of $\eta$ with respect to $x$ and $T$ is shown on Figure \ref{fig:visco}. Unlike other properties, the composition variation appears to be small. This behavior is different from other melts such as LiF-BeF$_2$ mixtures, for which viscosities larger by several orders of magnitudes are obtained on the BeF$_2$-rich side~\cite{salanne2007a}. This is due to the strong network-forming ability of BeF$_2$, which is not observed for ThF$_4$. The largest viscosities are thus obtained for the lower temperatures, around the eutectic composition.

\begin{figure}[ht!]
\begin{center}
 \includegraphics[width=\columnwidth]{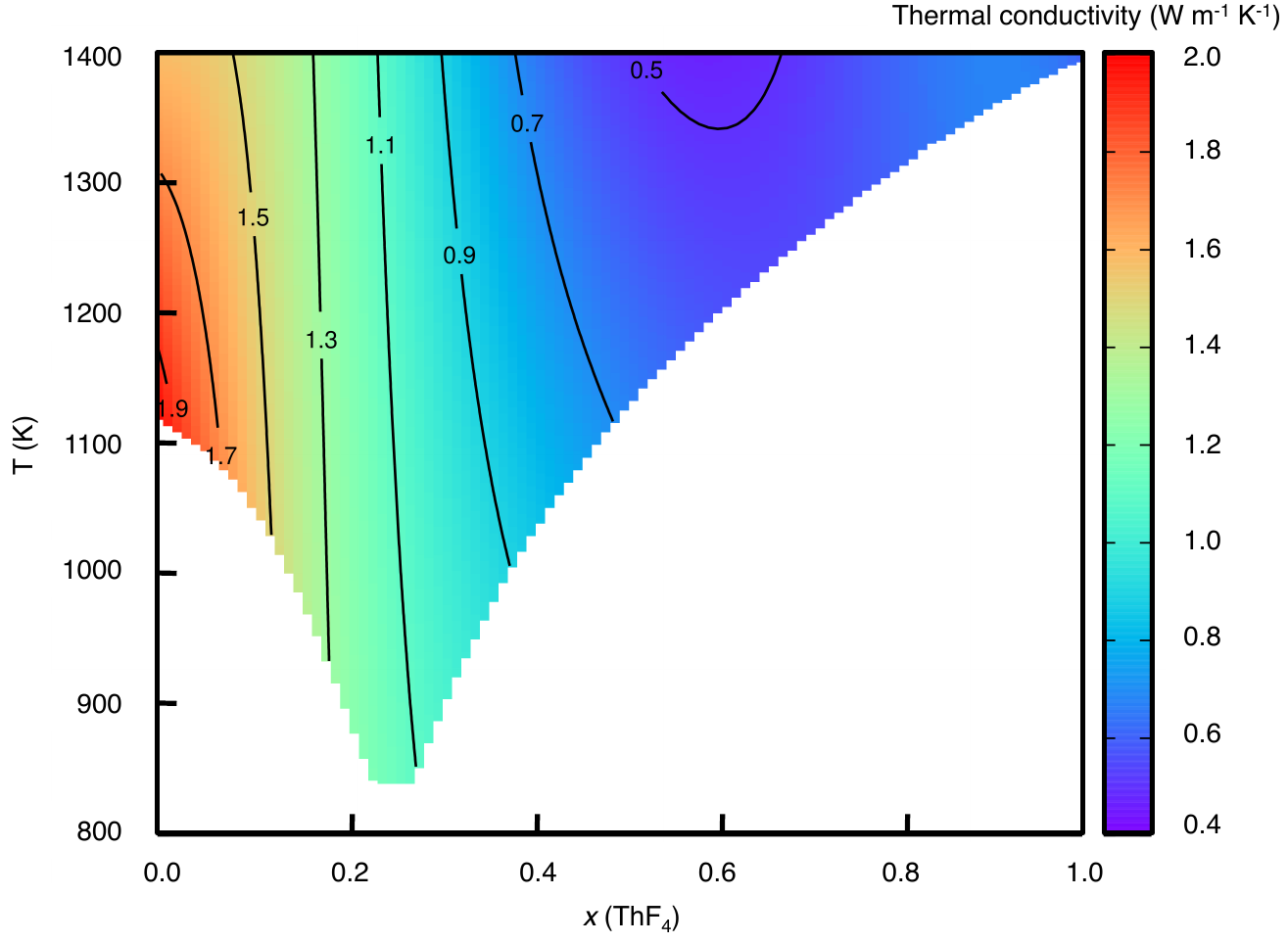}
\end{center}
\caption{Variation of the thermal conductivity of the LiF-ThF$_4$ melts with composition and temperature.  The white region corresponds to temperatures below the liquidus.}
\label{fig:thermal}
\end{figure}

Finally, the thermal conductivity is the most complex quantity to calculate. Like the viscosity, it is obtained using the Green-Kubo method~\cite{ohtori2009a}, but in that case several phenomelogical coefficients need to be calculated from the integration of the corresponding microscopic fluxes auto-correlation functions. Nevertheless, it was shown by Ohtori {\it et al.} that its temperature variation is very small, and is a consequence of the change of the density of the melt only~\cite{ohtori2009b}. We used the same analytic function as for the density, and the following expression was obtained.
\begin{eqnarray}
\lambda(x,T)&=&3.6014-0.0014497\times T \nonumber \\
            & &+(-24.140+0.017325\times T)\times x\nonumber \\
            & & +(87.337-0.072664\times T)\times x^2 \nonumber \\
            & &+(-126.47+0.10924\times T)\times x^3 \nonumber \\
            & & +(61.298-0.053182\times T)\times x^4,
\end{eqnarray}  
\noindent with $\chi(\lambda)$~=~5.9~\%, which is again well within the simulation error bars. The $\lambda(x,T)$ map is plotted on Figure \ref{fig:thermal}, interestingly we observe a minimum for a composition of $x$~=~0.6. Unlike the viscosity, the thermal conductivity of liquids does not show substantial variations when the structure of the melt changes (upon formation of a network, for example). Some efforts are currently made in order to improve our understanding of this quantity from the physical chemistry point of view in molten salts~\cite{ohtori2009b} as well as in molecular solvents such as water~\cite{roemer2012a,roemer2012b}.

\begin{figure}[ht!]
\begin{center}
 \includegraphics[width=\columnwidth]{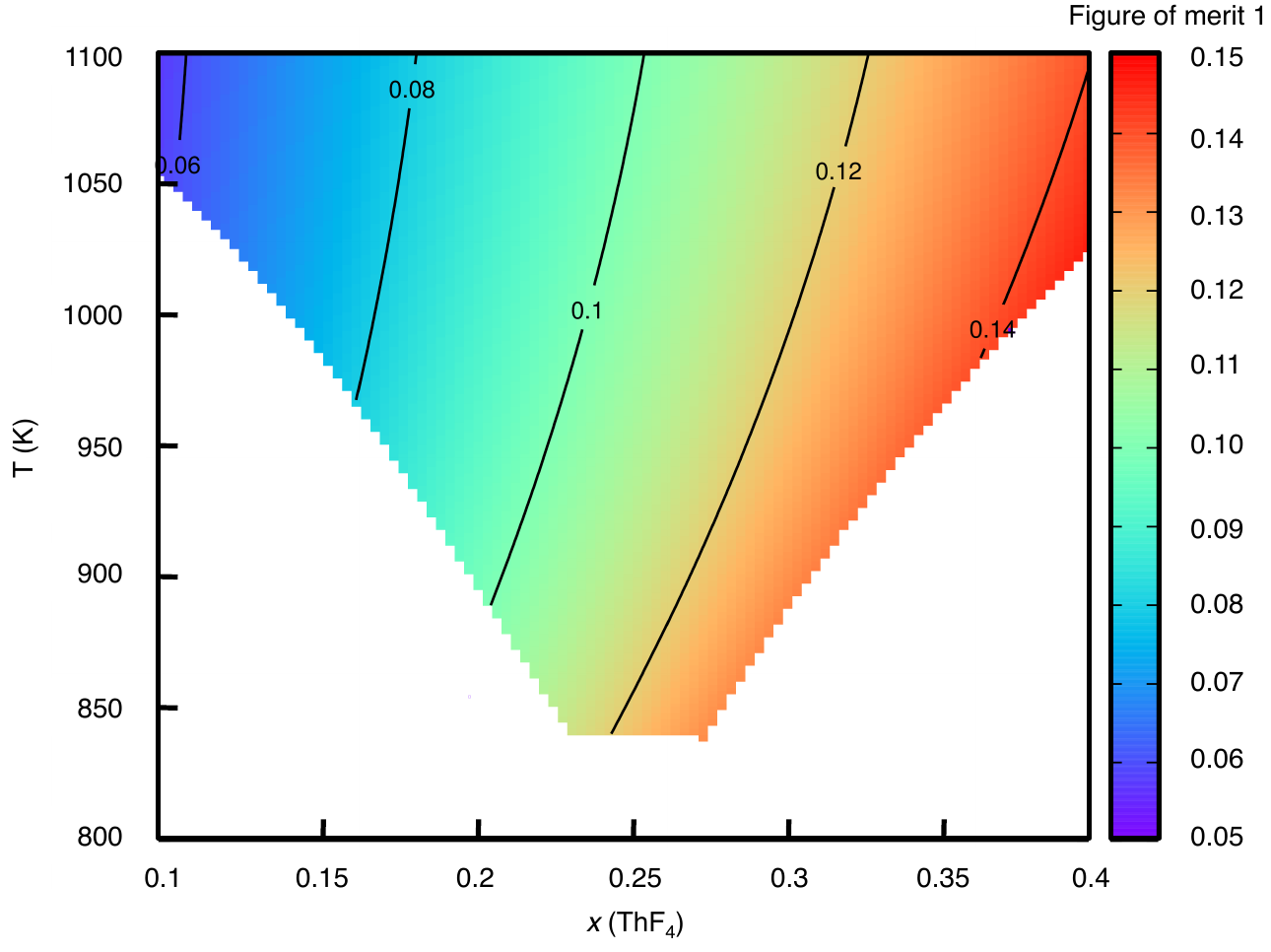}
\end{center}
\caption{Variation of the FOM1 (forced convection, turbulent r\'egime) of the LiF-ThF$_4$ melts with composition and temperature in the eutectic region.  The white region corresponds to temperatures below the liquidus.}
\label{fig:merit1}
\end{figure}

It is then possible to calculate the FOMs as defined by Equation \ref{eq:fomeq} and the sets of exponents given in Table \ref{tab:fom} at any thermodynamic point. The example of FOM1 in the eutectic region is shown on Figure \ref{fig:merit1}; similar plots are provided for FOM2, 3 and 4 in the supplementary information. The smaller the FOM is, the better heat-transfer performance. We note that the FOMs of LiF-ThF$_4$ are relatively similar to those of LiF-NaF-KF and NaF-ZrF$_4$ eutectic mixtures~\cite{salanne2009a}, which are among the candidates for being used in another molten salt-based reactor, the advanced high temperature reactor~\cite{forsberg2005a}. The figure reveals that a LiF-rich melt is favorable for the thermal properties. Since the thorium is the fertile material, from which the fissile $^{233}$U is formed during the reactor operation,  a compromise would have to be found for choosing an optimal composition for the reactor performance. Furthermore, one of the most important technological difficulties which needs to be overcome is to have container materials with a very high resistance to corrosion. A consequence of this is that the temperature has to be kept as low as possible, leading to composition choices close to the eutectic. The spatial variation of the temperature may be relatively large inside the reactor core. Here we observe that the FOMs tend to decrease with increasing temperature, which should ensure a good extraction of the heat at all the conditions. 

%\begin{figure}[ht!]
%\begin{center}
% \includegraphics[width=\columnwidth]{../Figures/merit2}
%\end{center}
%\caption{Variation of the FOM2 (natural convection, turbulent r\'egime) of the LiF-ThF$_4$ melts with composition and temperature.}
%\label{fig:merit2}
%\end{figure}

%\begin{figure}[ht!]
%\begin{center}
% \includegraphics[width=\columnwidth]{../Figures/merit3}
%\end{center}
%\caption{Variation of the FOM3 (natural convection, laminar r\'egime) of the LiF-ThF$_4$ melts with composition and temperature.}
%\label{fig:merit3}
%\end{figure}

\section{Conclusion}
The development of reliable interaction potentials, with {\it ab initio} accuracy, and the availability of large computer resources now allow us to build reliable thermophysical databases for molten salts. Such an approach is complementary to experimental measurements, yielding a large number of data at a very reasonable cost. In the present work we have illustrated this by calculating the density, the thermal expansion, the heat capacity, the viscosity and the thermal conductivity of LiF-ThF$_4$ mixtures for a wide range of compositions and temperatures. By fitting the calculated data with analytic functions, we have been able to map the variations of these properties for any thermodynamic point, and to determine the figures of merit which are necessary for designing molten salts reactor. 

Although LiF and ThF$_4$ are the main components of the MSFR, it is sure that small amounts of other fluorides (UF$_3$, UF$_4$ and PuF$_3$, mainly) will also be present. In addition a large number of species will be formed due to the fission and the neutron capture reactions, and it remains impossible to extend the present work for all the elements. Therefore only the most important ones can be considered. 
However this difficulty can be overcomed by using thermodynamic models, which have up to now been fitted on experimental data only. The improvement of the thermodynamic models by the inclusion of the large amount of data provided by MD simulations should allow for the extension of the range of their applicability in the future.

\section*{Acknowledgements}
This work was supported by the EVOL project in the 7$^{\rm th}$ Framework Programme of the European Commission (Grant agreement No. 249696) and by funding from the Natural
Sciences and Engineering Research Council of Canada (NSERC).Computational
resources for this work were provided by the R\'eseau Qu\'eb\'ecois de Calcul de
Haute Performance (RQCHP).

\bibliography{references}

\begin{thebibliography}{10}

\bibitem{locatelli2013a}
G.~Locatelli, M.~Mancini, and N.~Todeschini.
\newblock Generation IV nuclear reactors: Current status and future prospects.
\newblock {\em Energ. Policy}, 61:1503--1520, 2013.

\bibitem{mathieu2006a}
L.~Mathieu, D.~Heuer, R.~Brissot, C.~Garzenne, C.~{Le Brun}, D.~Lecarpentier,
  E.~Liatard, J.~M. Loiseaux, O.~Meplan, E.~Merle-Lucotte, A.~Nuttin, E.~Walle,
  and J.~Wilson.
\newblock The thorium molten salt reactor: moving on from the MSBR.
\newblock {\em Prog. Nucl. Energy}, 48:664--679, 2006.

\bibitem{aufiero2014a}
M.~Aufiero, M.~Brovchenko, A.~Cammi, I.~Clifford, O.~Geoffroy, D.~Heuer,
  A.~Laureau, M.~Losa, L.~Luzzi, E.~Merle-Lucotte, M.~E. Ricotti, and H.~Rouch.
\newblock Calculating the effective delayed neutron fraction in the molten salt
  fast reactor: Analytical, deterministic and monte carlo approaches.
\newblock {\em Ann. Nucl. Energy}, 65:78--90, 2014.

\bibitem{brovchenko2013a}
M.~Brovchenko, D.~Heuer, E.~Merle-Lucotte, M.~Allibert, V.~Ghetta, A.~Landreau,
  and P.~Rubiolo.
\newblock Design-related studies for the preliminary safety assessment of the
  molten salt fast reactor.
\newblock {\em Nucl. Sci. Eng.}, 175:329--339, 2013.

\bibitem{delpech2009a}
S.~Delpech, E.~Merle-Lucotte, D.~Heuer, M.~Allibert, V.~Ghetta, C.~Le-Brun,
  X.~Doligez, and G.~Picard.
\newblock Reactor physics and reprocessing scheme for innovative molten salt
  reactor system.
\newblock {\em J. Fluorine Chem.}, 130:11--17, 2009.

\bibitem{hamel2007a}
C.~Hamel, P.~Chamelot, A.~Laplace, E.~Walle, O.~Dugne, and P.~Taxil.
\newblock Reduction process of uranium (\uppercase{IV}) and uranium
  (\uppercase{III}) in molten fluorides.
\newblock {\em Electrochim. Acta}, 52:3995--4003, 2007.

\bibitem{gibilaro2009a}
M.~Gibilaro, L.~Massot, P.~Chamelot, L.~Cassayre, and P.~Taxil.
\newblock Electrochemical extraction of europium from molten fluoride media.
\newblock {\em Electrochim. Acta}, 55:281--287, 2009.

\bibitem{taxil2009a}
P.~Taxil, L.~Massot, C.~Nourry, M.~Gibilaro, P.~Chamelot, and L.~Cassayre.
\newblock Lanthanides extraction processes in molten fluoride media:
  Application to nuclear spent fuel reprocessing.
\newblock {\em J. Fluorine Chem.}, 130:94--101, 2009.

\bibitem{gibilaro2011a}
M.~Gibilaro, L.~Cassayre, O.~Lemoine, L.~Massot, O.~Dugne, R.~Malmbeck, and
  P.~Chamelot.
\newblock Direct electrochemical reduction of solid uranium oxide in molten
  fluoride salts.
\newblock {\em J. Nucl. Mater.}, 414:169--173, 2011.

\bibitem{benes2009b}
O.~Benes and R.~J.~M. Konings.
\newblock Thermodynamic properties and phase diagrams of fluoride salts for
  nuclear applications.
\newblock {\em J. Fluorine Chem.}, 130:22--29, 2009.

\bibitem{benes2013a}
O.~Benes and R.~J.~M. Konings.
\newblock Thermodynamic assessment of the LiF--CeF$_3$--ThF$_4$ system: Prediction of
  PuF$_3$ concentration in a molten salt reactor fuel.
\newblock {\em J. Nucl. Mater.}, 435:164--171, 2013.

\bibitem{capelli2013a}
E.~Capelli, O.~Benes, M.~Beilmann, and R.J.M. Konings.
\newblock Thermodynamic investigation of the LiF-ThF$_4$ system.
\newblock {\em J. Chem. Thermodyn.}, 58:110--116, 2013.

\bibitem{benes2012a}
O.~Benes and R.J.M. Konings.
\newblock Molten salt reactor fuel and coolant.
\newblock {\em Comprehensive Nuclear Materials}, 3:359--389, 2012.

\bibitem{benes2012b}
O.~Benes, R.~J.~M. Konings, D.~Sedmidubsk\'y, M.~Beilmann, O.~S. Valu,
  E.~Capelli, M.~Salanne, and S.~Nichenko.
\newblock A comprehensive study of the heat capacity of CsF from T = 5 K to T =
  1400 K.
\newblock {\em J. Chem. Thermodyn.}, 57:92--100, 2012.

\bibitem{khoklov2011a}
V.~Kohklov, I.~Korzun, V.~Dokutovich, and E.~Filatov.
\newblock Heat capacity and thermalconductivity of molten ternary lithium,
  sodium, potassium, and zirconium fluorides mixtures.
\newblock {\em J. Nucl. Mater.}, 410:32--38, 2011.

\bibitem{madden2006a}
P.~A. Madden, R.~J. Heaton, A.~Aguado, and S.~Jahn.
\newblock From first-principles to material properties.
\newblock {\em J. Mol. Struct.: THEOCHEM}, 771:9--18, 2006.

\bibitem{salanne2011c}
M.~Salanne and P.~A. Madden.
\newblock Polarization effects in ionic solids and melts.
\newblock {\em Mol. Phys.}, 109:2299--2315, 2011.

\bibitem{salanne2012b}
M.~Salanne, B.~Rotenberg, C.~Simon, S.~Jahn, R.~Vuilleumier, and P.~A. Madden.
\newblock Including many-body effects in models for ionic liquids.
\newblock {\em Theor. Chem. Acc.}, 131:1143, 2012.

\bibitem{heaton2006a}
R.~J.~Heaton, R. Brookes, P.~A.~Madden, M.~Salanne, C.~Simon, and P. Turq
\newblock A first-principles description of liquid \uppercase{B}e\uppercase{F}$_2$ and its mixtures with \uppercase{L}i\uppercase{F}: 1. Potential development and pure \uppercase{B}e\uppercase{F}$_2$.
\newblock {\em J. Phys. Chem. B}, 110:11454--11460, 2006.

\bibitem{pauvert2010a}
O.~Pauvert, D.~Zanghi, M.~Salanne, C.~Simon, A.~Rakhmatullin, H.~Matsuura,
  Y.~Okamoto, F.~Vivet, and C.~Bessada.
\newblock In situ experimental evidence for a nonmonotonous structural
  evolution with composition in the molten
  \uppercase{L}i\uppercase{F}-\uppercase{Z}r\uppercase{F}$_4$ system.
\newblock {\em J. Phys. Chem. B}, 114:6472, 2010.

\bibitem{levesque2013b}
M.~Levesque, V.~Sarou-Kanian, M.~Salanne, M.~Gobet, H.~Groult, C.~Bessada, and
  A.-L. Rollet.
\newblock Structure and dynamics in yttrium-based molten rare earth alkali
  fluorides.
\newblock {\em J. Chem. Phys.}, 138:184503, 2013.

\bibitem{dewan2013a}
L.~C. Dewan, C.~Simon, P.~A. Madden, L.~W. Hobbs, and M.~Salanne.
\newblock Molecular dynamics simulation of the thermodynamic and transport
  properties of the molten salt fast reactor fuel LiF--ThF$_4$.
\newblock {\em J. Nucl. Mater.}, 434:322--327, 2013.

\bibitem{tang1984a}
K.~T. Tang and J.~P. Toennies.
\newblock An improved simple model for the van der {W}aals potential based on
  universal damping functions for the dispersion coefficients.
\newblock {\em J. Chem. Phys.}, 80:3726--3741, 1984.

\bibitem{aguado2003a}
A.~Aguado and P.~A. Madden.
\newblock Ewald summation of electrostatic multipole interactions up to the
  quadrupolar level.
\newblock {\em J. Chem. Phys.}, 119:7471--7483, 2003.

\bibitem{bonilla1958a}
C.~F. Bonilla.
\newblock {\em Nuclear Engineering Handbook}, chapter 6.5, pages 9--90.
\newblock 1958.

\bibitem{heaton2008a}
R.~J. Heaton and P.~A. Madden.
\newblock Fluctuating ionic polarizabilities in the condensed phase:
  first-principles calculations of the \uppercase{R}aman spectra of ionic
  melts.
\newblock {\em Mol. Phys.}, 106:1703--1719, 2008.

\bibitem{salanne2006a}
M.~Salanne, C.~Simon, P.~Turq, R.~J. Heaton, and P.~A. Madden.
\newblock A first-principles description of liquid
  \uppercase{B}e\uppercase{F}$_2$ and its mixtures with
  \uppercase{L}i\uppercase{F}: 2. network formation in
  \uppercase{L}i\uppercase{F}-\uppercase{B}e\uppercase{F}$_2$.
\newblock {\em J. Phys. Chem. B}, 110:11461--11467, 2006.

\bibitem{saroukanian2009a}
V.~Sarou-Kanian, A.-L. Rollet, M.~Salanne, C.~Simon, C.~Bessada, and P.~A.
  Madden.
\newblock Diffusion coefficients and local structure in basic molten fluorides:
  {\it in situ} \uppercase{NMR} measurements and molecular dynamics
  simulations.
\newblock {\em Phys. Chem. Chem. Phys.}, 11:11501--11506, 2009.

\bibitem{ohtori2009a}
N.~Ohtori, M.~Salanne, and P.~A. Madden.
\newblock Calculations of the thermal conductivities of ionic materials by
  simulation with polarizable interaction potentials.
\newblock {\em J. Chem. Phys.}, 130:104507, 2009.

\bibitem{beilmann2013a}
M.~Beilmann, O.~Benes, E.~Capelli, V.~Reuscher, R.~J.~M. Konings, and Th.
  Fangh\"anel.
\newblock Excess heat capacity in liquid binary alkali-fluoride mixtures.
\newblock {\em Inorg. Chem.}, 52:2404--2411, 2013.

\bibitem{hansen-livre}
J.-P. Hansen and I.R. McDonald.
\newblock {\em Theory of simple liquids}.
\newblock Academic Press, 4th edition, 1986.

\bibitem{salanne2007a}
M.~Salanne, C.~Simon, P.~Turq, and P.~A. Madden.
\newblock Simulation of the liquid-vapor interface of molten
  \uppercase{L}i\uppercase{B}e\uppercase{F}$_3$.
\newblock {\em C. R. Chim.}, 10:1131--1136, 2007.

\bibitem{ohtori2009b}
N.~Ohtori, T.~Oono, and K.~Takase.
\newblock Thermal conductivity of molten alkali halides: temperature and
  density dependence.
\newblock {\em J. Chem. Phys.}, 130:044505, 2009.

\bibitem{roemer2012a}
F.~Roemer and F.~Bresme.
\newblock Heat conduction and thermomolecular orientation in diatomic fluids: a
  non-equilibrium molecular dynamics study.
\newblock {\em Mol. Simulat.}, 38:1198--1208, 2012.

\bibitem{roemer2012b}
F.~Roemer, A.~Lervik, and F.~Bresme.
\newblock Nonequilibrium molecular dynamics simulations of the thermal
  conductivity of water: A systematic investigation of the spc/e and tip4p/2005
  models.
\newblock {\em J. Chem. Phys.}, 137:074503, 2012.

\bibitem{salanne2009a}
M.~Salanne, C.~Simon, P.~Turq, and P.~A. Madden.
\newblock Heat-transport properties of molten fluorides: determination from
  first-principles.
\newblock {\em J. Fluorine Chem.}, 130:38--44, 2009.

\bibitem{forsberg2005a}
C.~Forsberg.
\newblock The advanced high-temperature reactor: high-temperature fuel, liquid
  salt coolant, liquid-metal-reactor plant.
\newblock {\em Progress in Nuclear Energy}, 47:32--43, 2005.

\end{thebibliography}

\end{document}